\documentclass[conference]{IEEEtran}
\usepackage{braket}
\usepackage{amsmath}
\usepackage{graphicx}
\usepackage{amssymb}
\usepackage{makecell}
\usepackage{tabularx}
\usepackage{array}
\usepackage{url}
\usepackage[table]{xcolor}
\usepackage{subfig}
\usepackage{afterpage}
\usepackage[bottom]{footmisc}
\usepackage[export]{adjustbox}
\graphicspath{ {./images/} }

\begin{document}
\title{
    Fault-Tolerant Error Detection above Break-Even for Multi-Qubit Gates
}
\author{
    \IEEEauthorblockN{Colburn Riffel, Reece Robertson, Peter Hendrickson}
    \IEEEauthorblockA{\textit{KBR, Inc.}\\
    Chantilly, VA USA \\
    Colburn.Riffel@us.kbr.com, Reece.Robertson@us.kbr.com, Pete.Hendrickson@us.kbr.com}
}

\maketitle
%-------------------------------------------
% Abstract
%-------------------------------------------
\begin{abstract}
    A fully fault-tolerant implementation of the quantum error-detecting Iceberg $[[2m, 2m-2, 2]]$ code applied to a Toffoli circuit achieved beyond-break-even error detection on a leading trapped-ion quantum computer, where the effect of encoding a circuit with a quantum error-detection code enables increased fidelity compared to an unencoded circuit.
    This code was also applied to Bell state preparation circuits, where a lean non-fault-tolerant implementation of the Iceberg code enables a fidelity gain as well. 
    This highlights the important point that, at least for small-scale circuits with a substantial portion of error-free runs, it can be effective simply to use error detection to filter out the runs with errors. 
    Furthermore, experiments performed in this work highlight the necessity for judicious compilation of circuits not only for a given hardware but also within a quantum error detection code.
\end{abstract}


%-------------------------------------------
% Paper Body
%-------------------------------------------
\section{Introduction}\label{introduction}
Quantum computing has produced algorithms to solve many important problems that achieve a speedup compared to their classical counterparts. These algorithms include Shor's algorithm for factoring~\cite{Shor1997}, the Harrow, Hassidim, and Lloyd (HHL) algorithm for solving linear systems of equations~\cite{Harrow2009}, and Grover's algorithm for rapid search of unstructured data~\cite{Grover1996}. However, one of the biggest obstacles to achieving the theoretical speedups of these algorithms on current noisy, intermediate-scale quantum (NISQ) hardware is the presence of noise within the quantum computers \cite{Robertson2025Simons, Robertson2026Gibbs}.

Efforts have been made to scale noiseless quantum algorithm simulations \cite{Orus2019, Robertson2025Unique}, yet it remains important to pursue noise reduction on real quantum hardware. To this end, several different methods for combating noise in a quantum computer have been developed. These include quantum error mitigation (QEM), quantum error detection (QED), and quantum error correction (QEC).

First, QEM aims to reduce the impact of noise on the overall results of the circuit, but does not correct or detect errors on an individual shot (or run of the circuit). Notable QEM techniques include dynamical decoupling~\cite{Khodjasteh2005}, zero noise extrapolation~\cite{Van2023}, and probabilistic error cancellation~\cite{Giurgica2020}.

Second, QED accounts for errors on each shot of the circuit by detecting and reporting them. Shots that yield a reported error may be rejected so that results are gathered from the clean shots where no errors were observed. QED codes do not attempt to correct errors. A significant QED code is the Iceberg $[[2m, 2m-2, 2]]$ code~\cite{Chao_2018, Self_2024}.

Third, QEC improves the fidelity of the circuit by detecting and correcting errors on each shot of the circuit. Notable QEC codes include Shor's QEC code~\cite{shor1995scheme}, Steane's QEC code~\cite{steane1996error}, topological codes~\cite{kitaev2003fault}, and quantum low-density parity check (qLDPC) codes~\cite{breuckmann2021quantum}. QEC requires a higher gate overhead than QED, but QED attains the comparative simplicity of its circuits at the cost of requiring more shots for each experiment. 

The largest obstacle to implementing QED and QEC codes is the increased overhead in the number of gates needed to execute a circuit.\footnote{As opposed to performing a circuit, QEC can be used to extend how long a logical state remains idle without being corrupted. This is called the \textit{quantum memory problem}~\cite{terhal2015quantum}. Only QEC codes can be used for this purpose, and, as this paper is focused on QED, the quantum memory problem will not be discussed further in this paper.} Although the purpose of applying QED or QEC to a quantum algorithm is to increase the circuit fidelity, on NISQ hardware using more physical gates to perform the circuit introduces more noise, often resulting in lower net fidelity \cite{Riffel2025}. To combat this issue, fault-tolerant operations have been developed. \textit{Fault-tolerance}, simply put, ensures that if an error occurs, it does not propagate into more errors than the code can handle. However, this introduces an additional increase in the number of physical gates required beyond the basic overhead for the code~\cite{gottesman1997stabilizer}. The point at which performing QED or QEC results in a net even fidelity is referred to as the \textit{break-even point}. If a circuit achieves above break-even fidelity, the effect of encoding a circuit with a QED or QEC code increases the fidelity compared to an unencoded circuit~\cite{sivak2023real}.

There have been several successful attempts at achieving fidelities above the break-even point on real quantum hardware. Experiments using QEC codes to achieve fidelity gain have been performed on Google's Willow chip~\cite{google2025quantum}, Quantinuum's H2 processors~\cite{paetznick2024demonstration, reichardt2024demonstration, hong2024entangling}, and an IBM Falcon processor~\cite{gupta2024encoding}. For QED, there have been a few successful attempts at achieving above break-even fidelity using: QAOA circuits~\cite{jin2025Icebergtipcocompilationquantum}, the Deutsch-Jozsa algorithm~\cite{Singh2025}, and mirror circuits~\cite{Self_2024}. Other experiments have demonstrated fault-tolerant entangling gates encoded with color codes \cite{Postler2022, Anderson2022, Postler2024, Kim2025}. In particular, Menendez et al. \cite{Menendez2024} demonstrated an advantage from the $[[8,3,2]]$ color code when executing non-Clifford $CCZ$ gates on an 11-qubit trapped-ion computer. In this paper, we use the more compact $[[2m,2m-2,2]]$ Iceberg code to demonstrate fidelity gain from QED when executing both Clifford $CX$ and non-Clifford Toffoli ($CCX$) circuits on a 36-qubit trapped-ion computer. This is significant, as the Toffoli gate is a subroutine in many important quantum algorithms, it is sufficient for universal computation \cite{aharonov2003simpleprooftoffolihadamard}, and it is a three-qubit logical operation that on current hardware is decomposed into one- and two-qubit gate operations \cite{Nielsen2010}.

This paper aims to analyze the Iceberg code when performing both fault-tolerant and non-fault-tolerant multi-qubit Toffoli gates. We implemented the Toffoli gate on a trapped-ion quantum computer using an Iceberg code both with and without fault-tolerance. We also apply the Iceberg code to Bell state circuits, to reveal an indication of how the performance of the Iceberg code scales as a function of the complexity of the original circuit.

Background on the necessary QED theory is given in Section \ref{background}. Precise details of all circuit implementations used in this study are given in Section \ref{methods}. We compared both fault-tolerant and non-fault-tolerant Iceberg circuits to each other, and to an unencoded baseline circuit. As shown in Section \ref{results}, fault-tolerant circuits successfully provided increased fidelity compared to non-fault-tolerant circuits. Moreover, fault-tolerant circuits were shown to achieve better than break-even fidelity for specific inputs to the Toffoli gate. The experiments performed in this work highlight the necessity for judicious compilation of circuits not only for a given hardware but also within a QED or QEC code. Concluding remarks drawn from these results are given in Section \ref{conclusions}.

\section{Quantum Error Detection and Correction}\label{background}
This background section provides a brief overview of the relevant context of quantum error detection (and correction) in general and the Iceberg code in particular. In this section, we review the foundational material introduced by Gottesman~\cite{gottesman1997stabilizer} along with the more recent work on the Iceberg code by Chao and Reichardt~\cite{Chao_2018} and Self et al.~\cite{Self_2024}. Excellent introductions to this field include~\cite{terhal2015quantum, Nielsen2010, gottesman2007fault, gottesman2010introduction, lidar2013quantum}. We will summarize the necessary concepts from these sources to ensure a solid foundation and will highlight specific aspects.

\subsection{Quantum Error Detection and Correction}

In general, a QED or QEC code encodes $k$ logical qubits into $n$ physical qubits by mapping states onto codewords. These codewords are specified by the code. A code is denoted $[[n,k,d]]$, where $n$ is the number of physical qubits, $k$ is the number of logical qubits, and $d$ is the distance of the code. The \textit{distance} of the code is the minimum weight of a Pauli operator to transform one codeword into another. Double brackets are used to denote that a code is a quantum error correction code and distinguish it from a classical $[n,k,d]$ code~\cite{lidar2013quantum}.

For QEC codes, the number of errors that the code can correct ($t$) is
\begin{equation}
    t = \left\lfloor\frac{d - 1}{2}\right\rfloor.
\end{equation}
Similarly, a QED code can detect $d-1$ errors~\cite{gottesman2010introduction}. Therefore, a code with distance 2 can only detect a single error, but it cannot correct the error. Hence, this code can be used for QED but not QEC.

Both QED and QEC codes begin by encoding \textit{blocks} of logical qubits from multiple physical qubits. After the encoding procedure, logical gates are applied to the encoded blocks. These gates perform the same logical operation on encoded qubits as the analogous physical operation performs on unencoded qubits~\cite{gottesman2007fault}. As will be discussed further in Section \ref{fault-tolerance}, these gates can be, but need not be, fault-tolerant.

After encoding and applying logical operations, errors are detected via syndrome measurement. \textit{Syndromes} are specific observables that can be measured to indicate the presence of errors. In the case of QEC, syndromes also indicate how to correct any error that may have occured.\footnote{A subtlety to note here is that the syndrome will indicate the method to correct the error which \textit{most likely} occured.} The syndrome measurement is performed with \textit{ancilla qubits} (extra qubits outside of an encoded block) in order to preserve the logical state of the code block and not collapse it with measurements.

Lastly, the code is either measured to determine a logical output or decoded for some further use, such as other computations~\cite{Nielsen2010}.

Many error correction and detection codes are \textit{stabilizer codes}. These codes have a compact expression using a specific group structure known as the \textit{stabilizer formalism}. In this formalism, a stabilizer group $\mathcal{S}$ is an Abelian subgroup of the Pauli group $P_n$ where
\begin{equation}
    P_n = \langle \pm I, \pm iI, \pm X, \pm iX, \pm Y, \pm iY, \pm Z, \pm iZ \rangle,
\end{equation}
with the exception that $-I \not\in \mathcal{S}$. The stabilizers of a code commute with all of the codewords but not the error states. Thus, when measured, they can be used to detect the presence of an error~\cite{terhal2015quantum}.

\subsection{Iceberg $[[2m,2m - 2,2]]$ Code}\label{iceberg-code}

The Iceberg $[[2m,2m - 2,2]]$ QED code can detect one error on any qubit since it has a distance $d=2$. Moreover, since $t=0$ for this code, it cannot correct any errors.

At a high level, the Iceberg code is implemented as follows. First, we take a physical circuit and add two qubits: a qubit on the top to track the $X$-parity (qubit $t$ for \textit{top}), and a qubit on the bottom to track the $Z$-parity (qubit $b$ for \textit{bottom}). We then transform our physical operations into logical operations that are constructed to track any parity changes that they introduce. Finally, we (optionally) add extra gates and ancillas as needed to ensure that all operations are fault-tolerant (see~\cite{Chao_2018} for examples of fault-tolerant and non-fault-tolerant operations). 

The Iceberg code is a stabilizer code, and the stabilizers of the code are
\begin{equation}
    \begin{aligned}
        S_x = \bigotimes_{j \in n} X_j,\\
        S_z = \bigotimes_{j \in n} Z_j.
    \end{aligned}
\end{equation}
The Iceberg code is encoded by creating a Greenberger-Horne-Zeilinger (GHZ) state \cite{Greenberger1989} over the appropriate number of physical qubits. Logical $X$ and $Z$ gates can be applied by performing
\begin{equation}
    \begin{aligned}
        \overline{X_i} = X_t X_i,\\
        \overline{Z_i} = Z_i Z_b,
    \end{aligned}
    \label{x-and-z}
\end{equation}
where the bar represents that the operation is a logical operation and the subscripts $t$ and $b$ indicate the top and bottom qubits, respectively~\cite{Self_2024}.

\begin{figure}
    \centering
    \subfloat[Targeted $H$ Gate]{
        \includegraphics[width=0.45\linewidth]{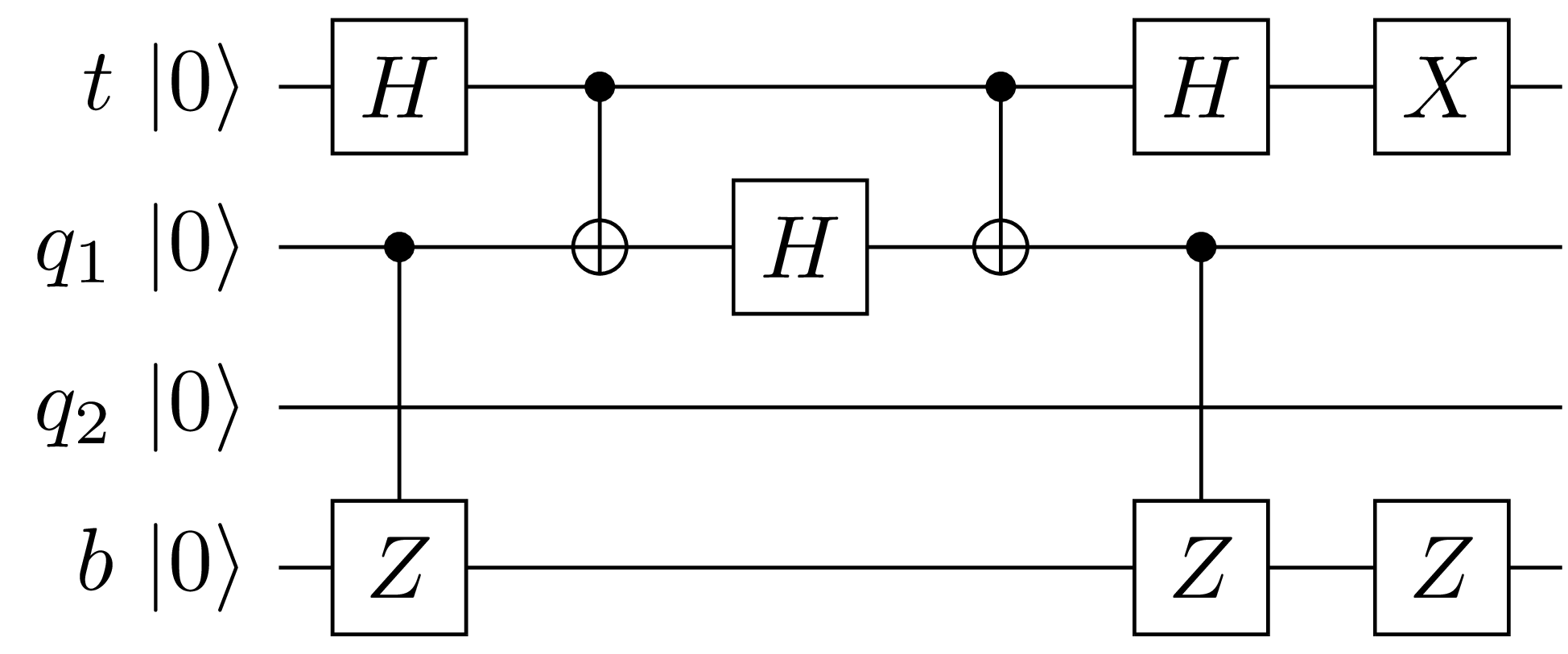}
        \label{fig:h-gate}
    }\\
    \subfloat[$CCZ$ Gate]{
        \includegraphics[width=0.65\linewidth]{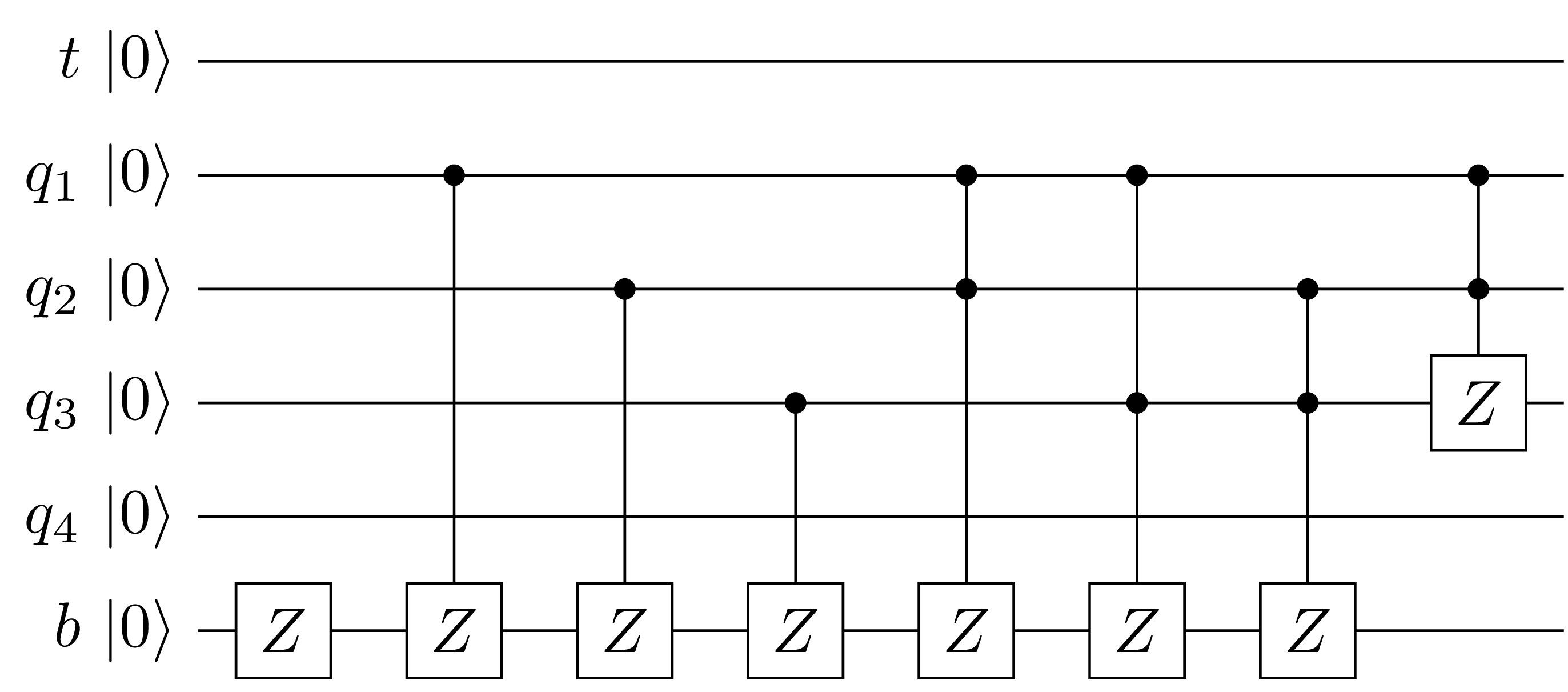}
        \label{fig:ccz-gate}
    }
    \caption{Circuit diagrams for Iceberg QED code logical operations. In Fig.~\ref{fig:h-gate}, a logical $H$ is applied to encoded qubit $q_1$. In Fig.~\ref{fig:ccz-gate}, a logical $CCZ$ gate has controls on encoded qubits $q_1$ and $q_2$, and a target on encoded qubit $q_3$. Note that the since the Iceberg code requires an even number of encoded qubits, the bottom encoded qubit is unused in both of these circuits.}
    \label{non-ft-Iceberg-ops}
\end{figure}

In this work, we used a logical gate set consisting of $H$ (Hadamard) and $CCZ$ (Toffoli\footnote{Surrounding the $X$ of a $CCX$ with $H$ gates transforms it into a $CCZ$, hence these gates are equivalent up to a basis change on the target qubit \cite{Mermin}.}) gates as implemented in~\cite{Chao_2018}. There are two different methods for applying the $H$ gate; as a targeted operation and as a transversal operation. The targeted $H$ applies an $H$ gate to a single qubit within the encoded block.\footnote{Note that the targeted $H$ gate is not a controlled-$H$, or $CH$, gate. Rather, it is an $H$ gate applied to (targeting) a single qubit within the encoded block.} The transversal $H$ will be described in Section \ref{fault-tolerance}. These gates are sufficient for universal quantum computation~\cite{aharonov2003simpleprooftoffolihadamard, shi2002both}.\footnote{Other gate combinations that yield universal computation are also possible.} The circuit diagrams for these gates are shown in Fig.~\ref{non-ft-Iceberg-ops}

Syndrome measurement for the Iceberg code can be performed by measuring both the $X$ stabilizer on an ancilla qubit and then measuring the $Z$ stabilizer in a similar manner. The $Z$ stabilizer can also be measured without an extra ancilla by verifying the parity of the final output was even (an odd parity on the final measurement denotes an error)~\cite{Self_2024}.

\subsection{Fault-Tolerant Operations}\label{fault-tolerance}

A \textit{fault-tolerant} operation is defined to be an operation where a single error will introduce, at most, one error in the logical block(s) on which the operation acts~\cite{gottesman1997stabilizer}. If our code can detect (or correct) a single error, the syndrome measurement can be applied after every operation, and the code will detect and reject (or correct) the error instead of propagating it into more errors later in the circuit. For example, the $X$ and $Z$ gates of \eqref{x-and-z} are fault-tolerant by construction.

In order to ensure operations are fault-tolerant, subroutines called gadgets are used. \textit{Gadgets} are composite operations representing a step in the circuit. They can represent any part of the circuit, including encoding, syndrome measurement, or a logical operation. Gadgets are useful because they mitigate the potential accumulation of errors through subsequent steps of the circuit. When the number of errors at physical locations in the gadget does not exceed a critical threshold during the application of the gadget, then the errors after applying the gadget will not be more than the quantum error correction or detection code can handle.~\cite{gottesman2010introduction}. This allows one to implement a gadget within a larger logical gate to ensure the overall logical gate maintains fault-tolerance.

\begin{figure}[b]
    \vspace{-20pt}
    \centering
    \subfloat[Fault-Tolerant $CZ$ Gadget]{
        \includegraphics[width=0.45\linewidth]{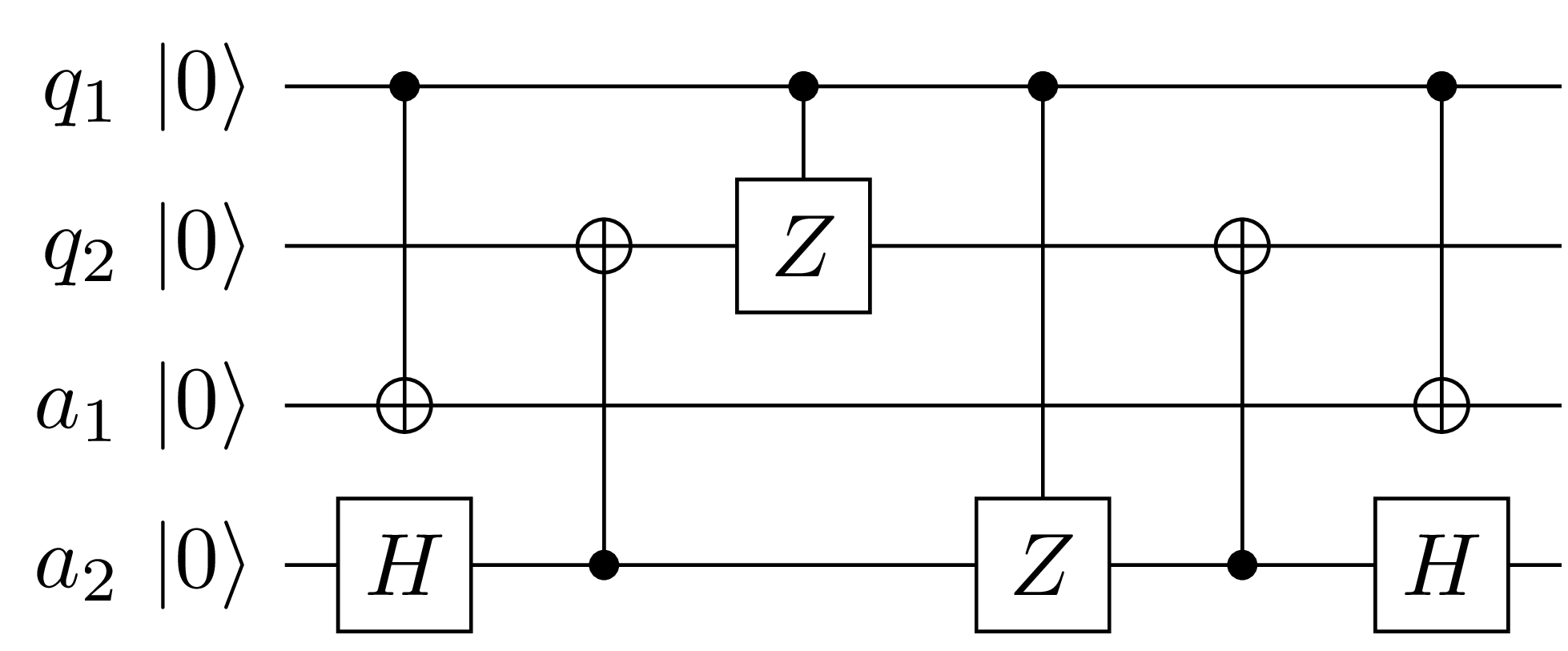}
        \label{cz-gadget}
    }\\
    \subfloat[Fault-Tolerant $CCZ$ Gadget]{
        \includegraphics[width=0.75\linewidth]{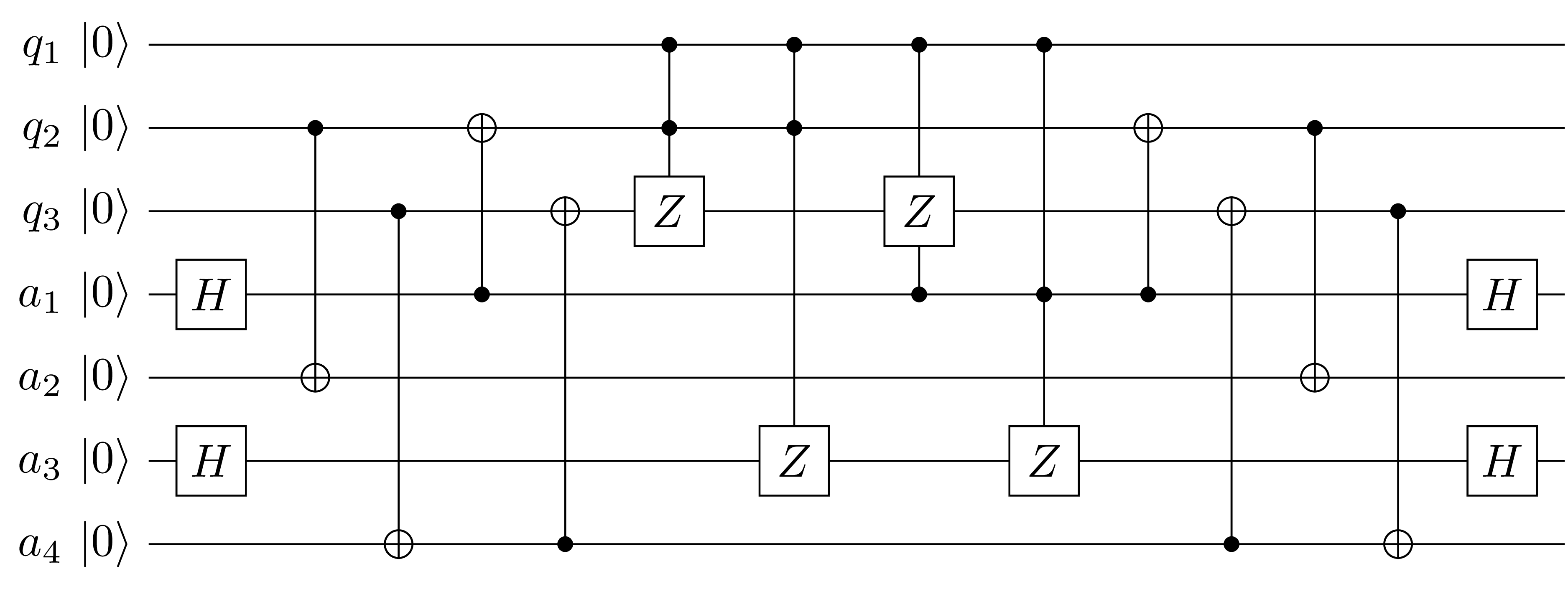}
        \label{ccz-gadget}
    }
    \caption{Circuit diagrams for fault-tolerant Iceberg code gadgets. The logical $CCZ$ gate in Fig.~\ref{non-ft-Iceberg-ops} can be made fault-tolerant by replacing each physical $CZ$ gate with the $CZ$ gadget in Fig.~\ref{cz-gadget} and each physical $CCZ$ gate with the $CCZ$ gadget in Fig.~\ref{ccz-gadget}.}
    \label{ft-Iceberg-ccz}
\end{figure}

\begin{figure*}[t]
    \centering
    \subfloat[Fault-Tolerant Targeted $H$ Gate]{\includegraphics[width=0.6\linewidth]{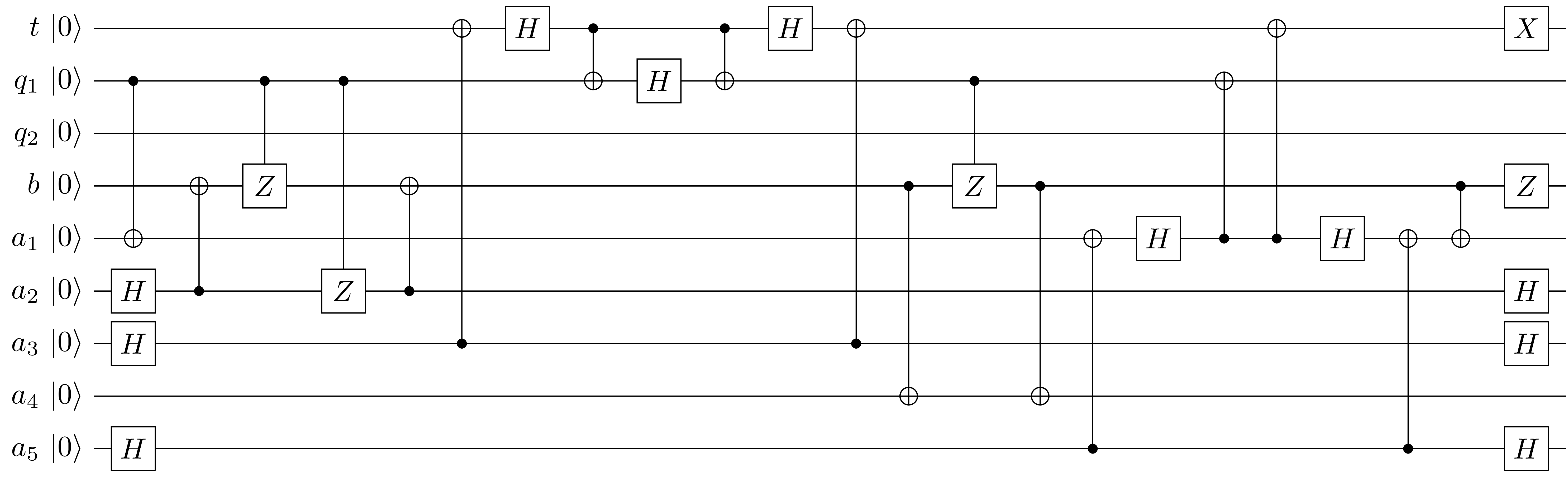}}\\
    \caption{Circuit diagram for the fault-tolerant targeted $H$ gate. The logical $H$ gate is applied to qubit $q_1$. Five ancilla qubits are required to ensure fault-tolerance, although with mid-circuit measurement only two ancillas are needed.}
    \label{ft-Iceberg-h}
\end{figure*}

For the Iceberg code, the encoding step is made fault-tolerant through a parity check onto one ancilla qubit~\cite{Self_2024}. For the $CCZ$ gate, fault-tolerance is achieved by replacing each physical $CZ$ and $CCZ$ gate in Fig.\ref{fig:ccz-gate} with the corresponding gadget of Fig.~\ref{ft-Iceberg-ccz}. The circuit for the fault-tolerant targeted $H$ gate is given in Fig.~\ref{ft-Iceberg-h}. These changes to the $H$ and $CCZ$ gates require the addition of five\footnote{With midcircuit measurement, the fault-tolerant targeted $H$ gate needs only two ancilla qubits.} or four ancilla qubits, respectively~\cite{Chao_2018}. The Iceberg code also has a unique property that applying an $H$ gate to every qubit results in a logical $H$ on every qubit~\cite{Chao_2018}. This is an example of a \textit{transversal gate}, which is fault-tolerant by construction~\cite{Nielsen2010}.

Finally, we conclude this section with one important observation: the complexity of a physical operation does not always correlate with the complexity of a logical operation. For example, the transversal $H$ gate that applies a logical $H$ to every qubit requires fewer operations than a targeted $H$ that applies a logical $H$ to a single qubit within the Iceberg code.

\section{Methods}\label{methods}
This paper follows the work of~\cite{Riffel2025} in analyzing the fidelity of the Iceberg $[[2m,2m - 2,2]]$ code by performing Toffoli and Bell state circuits within the code. Specifically, circuits encoded by the Iceberg code fault-tolerantly were compared to corresponding circuits encoded non-fault-tolerantly. Both of these configurations were benchmarked against a corresponding unencoded baseline circuit. 

We now consider the circuits on which the codes were applied. In order to test the efficacy of the code, five circuits were selected. Three circuits incorporated a Toffoli gate, while two others produced Bell states.

\begin{figure}[b]
    \vspace{-20pt}
    \centering
    \subfloat[
        \centering``Bell State'' \& ``Transversal-$H$ Bell State''
    ]{
        \includegraphics[width=0.25\linewidth, valign=m]{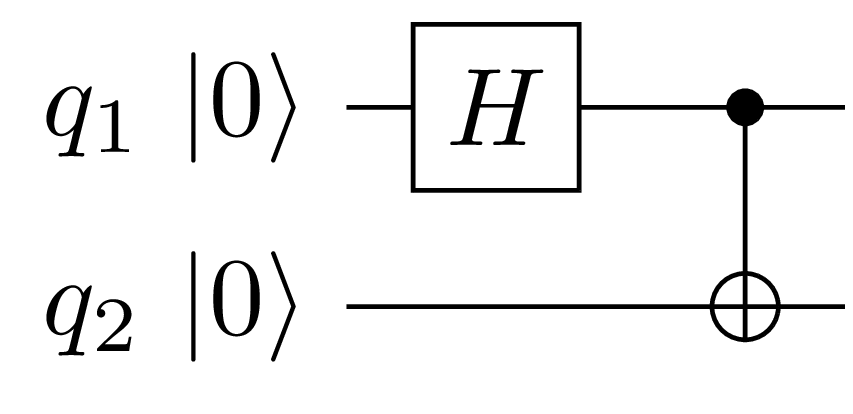}
        \label{bell-state}
    }\hfill
    \subfloat[\centering``$XX$ Toffoli''
    ]{
        \includegraphics[width=0.25\linewidth, valign=m]{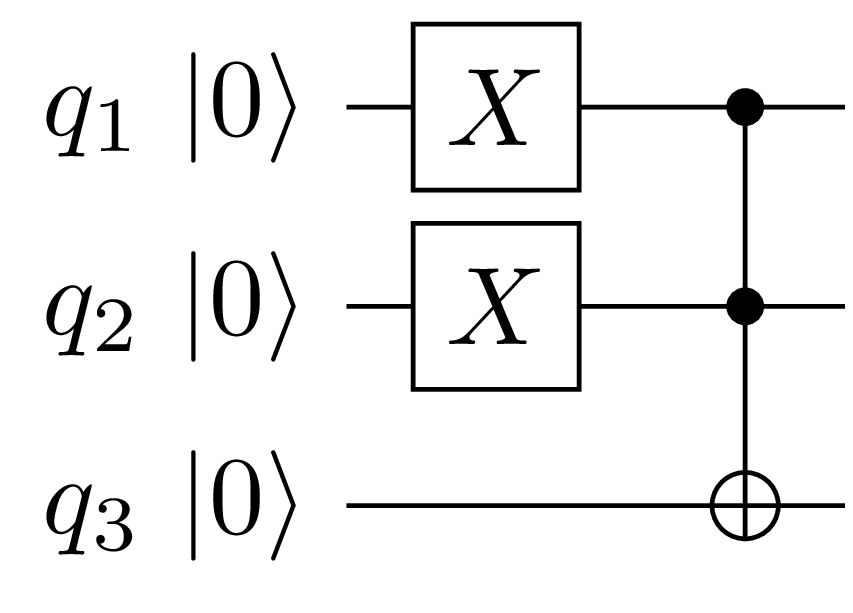}
        \label{xx-toffoli}
    }\hfill
    \subfloat[
        \centering``$HH$ Toffoli'' \& ``Transversal-$H$ Toffoli''
    ]{
        \includegraphics[width=0.25\linewidth, valign=m]{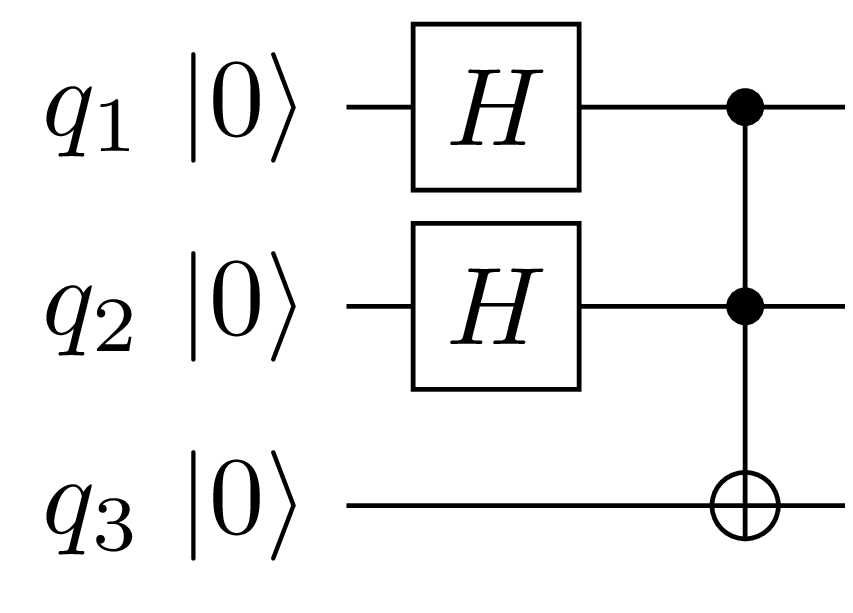}
        \label{hh-toffoli}
    }
    \caption{Unencoded configurations for the five circuits examined in this study. Each of these circuits was executed on a 36-qubit trapped-ion hardware device as shown (after compilation into hardware native gates), after Iceberg QED encoding with non-fault-tolerant operations, and after Iceberg QED encoding with fault-tolerant operations.}
    \label{logical-circuits}
\end{figure}

Both Bell state circuits create the $\ket{00}_L + \ket{11}_L$ state, where the subscript $L$ indicates a logical state. The physical circuit to produce this state is given in Fig.~\ref{bell-state}. The difference between the two logical Bell circuits used in this study arose from the method with which the $H$ gates were encoded. The first circuit performed a targeted $H$ for all $H$ gates. This circuit was named the ``Bell State'' circuit. The second circuit replaced the initial targeted $H$ gates\footnote{Since we have a gadget for the fault-tolerant $CZ$ operation rather than the $CX$ directly, we create a $CX$ by surrounding the $Z$ of a $CZ$ with $H$ gates. Hence, when represented with a $CZ$, the circuit in Fig.~\ref{bell-state} includes three $H$ gates, two of which precede the two-qubit gate.} with a transversal $H$ operation, and hence was named the ``Transversal-$H$ Bell State'' circuit.

Of the three circuits that involved Toffoli gates, one prepared the state $\ket{110}_L$ before applying the Toffoli gate (Fig.~\ref{xx-toffoli}), and the other two prepared the state $\ket{000}_L + \ket{110}_L$ before the Toffoli (Fig.~\ref{hh-toffoli}). The first circuit was deemed the ``$XX$ Toffoli'' circuit, and was included in this experiment to validate the action of the encoded Toffoli gate. The first circuit using the $\ket{000}_L + \ket{110}_L$ state implemented each $H$ via a logical targeted $H$ gate, and was named the ``$HH$ Toffoli'' circuit. The third circuit differed from the second only through the use of a transversal $H$ operation for the state preparation, and hence was named the ``Transversal-$H$ Toffoli'' circuit. These two circuits were included in the experiment to test the action of the encoded Toffoli gate on an entangled state.

Circuits were executed under three different \textit{configurations}. The first configuration was an unencoded baseline\footnote{That is, the circuits shown in Fig.~\ref{logical-circuits}, reexpressed in terms of the hardware native gate sets. In practice, on NISQ hardware this means that Toffoli gates are decomposed into a sequence of one- and two-qubit gates.} to assess hardware fidelity in the absence of error-detection. The second configuration was encoded with the Iceberg scheme using non-fault-tolerant operations (Fig.~\ref{non-ft-Iceberg-ops}). However, circuit encoding and syndrome readout were still performed fault-tolerantly. Lastly, the third configuration was encoded with the Iceberg scheme using fault-tolerant operations (Figs.~\ref{ft-Iceberg-ccz} \& \ref{ft-Iceberg-h}), with some caveats explained in the next paragraph. One \textit{run} of the circuit constitutes the execution of each of these configurations on the hardware, which allows for a comparison of fidelity across the different configurations.

First, for the ``$XX$ Toffoli'' circuit, it was necessary to sacrifice fault-tolerance of at least one operation in order to fit the encoded circuit on a trapped-ion device with 36 qubits and without mid-circuit measurement support. As such, the first logical targeted $H$ gate was performed without fault-tolerance.\footnote{This gate was chosen because errors introduced by this gate could trigger error flags in subsequent gates. Hence, by performing the first $H$ gate without fault-tolerance, we maximized the opportunity for any undetected errors to be caught by other checks in the algorithm.} The final encoded circuit required 6 ancillas and 30 working qubits for a total of 36 qubits, which fit the hardware constraints exactly.

The same problem faced the ``$HH$ Toffoli'' circuit, thus for the same reason we performed the first three logical targeted $H$ gates non-fault-tolerantly. This circuit also used 6 ancillas and 30 working qubits for a total of 36 qubits. However, for the ``Transversal-$H$ Toffoli'' circuit, the transversal Hadamard gate was fault-tolerant by construction. As such, we were able to perform all operations fault-tolerantly in this circuit, while again using 6 ancilla qubits and 30 working qubits. This highlights an important point---two different logical encoded implementations of the same physical unencoded circuit have different complexities. In other words, logical circuits should be carefully compiled within a QED or QEC code.

Once the three versions of each circuit (unencoded, non-fault-tolerant, and fault-tolerant) were obtained, each circuit was run with 2056 shots per run at three different times over the course of one calibration cycle: shortly after calibration, in between recalibrations, and shortly before recalibration. This allowed us to average over fluctuating fidelities within the device's recalibration cycle. Once the measurement results were obtained, the number of successful shots was used to calculate and compare fidelities for each version of the circuit. In the case encoded circuits, all shots that yielded an error were rejected prior to the fidelity calculation. The fidelities and rejection rates for encoded circuits are reported in the next section.

\section{Results}\label{results}
The results of experiments will be discussed in the following order: first the ``Bell State'' circuit (Table \ref{tab:bell-state}), next the ``Transversal-$H$ Bell State'' circuit (Table \ref{tab:transversal-bell-state}), then the ``$XX$ Toffoli'' circuit (Table \ref{tab:xx-toffoli}), followed by the ``$HH$ Toffoli'' circuit (Table \ref{tab:hh-toffoli}), and finally with the ``Transversal-$H$ Toffoli'' circuit (Table \ref{tab:transversal-toffoli}).

A table is dedicated to the results of each circuit. For each run a first row compares the unencoded and non-fault-tolerant configurations. The second row compares the unencoded and fault-tolerant configurations. The run number is given in the first column, the second column denotes whether the run was fault-tolerant, the third column gives the fidelity of the unencoded baseline, the fourth column gives the result of the encoded configuration, and the fifth reports the percentage of shots rejected by the encoded configuration due to a detected error. Runs where the encoded configuration surpassed the unencoded baseline--the break-even point--are highlighted in tables \ref{tab:bell-state}, \ref{tab:transversal-bell-state}, \ref{tab:xx-toffoli}, and \ref{tab:transversal-toffoli}.

\begin{table}[tb]
    \centering
    \caption{\textsc{Fidelity of ``Bell State'' Circuits}}
    \begin{tabular}{|c|c|c|c|c|}
        \hline
        \makecell{Run\\ Number} & \makecell{Fault\\ Tolerance} & \makecell{Unencoded\\ Baseline} & \makecell{Encoded\\ Fidelity} & \makecell{Detected\\ Errors} \\
        \hline
        \hline
        \rowcolor{blue!20}
        Run 1 & No & 99.07\% & 99.45\% & 11.18\% \\
        \hline
        Run 1 & Yes & 99.07\% & 98.75\% & 37.30\% \\
        \hline
        Run 2 & No & 99.12\% & 98.86\% & 10.06\% \\
        \hline
        Run 2 & Yes & 99.12\% & 98.81\% & 34.47\% \\
        \hline
        \rowcolor{blue!20}
        Run 3 & No & 98.93\% & 99.02\% & 10.25\% \\
        \hline
        Run 3 & Yes & 98.93\% & 98.69\% & 33.01\% \\
        \hline
    \end{tabular}
    \label{tab:bell-state}
    \vspace{-10pt}
\end{table}

For the ``Bell State'' circuit, performance surpassed the break-even point in two of the runs. Perhaps surprisingly, beyond break-even performance was observed with the \textit{non-fault-tolerant} configuration. However, due to the simplicity of the circuit, fidelity was very high for the unencoded runs of the ``Bell State'' circuit, hence there was little room for improvement. Moreover, the absolute difference in fidelity across all runs of this circuit is less than 1\%, so perceived successes and failures in surpassing the break-even point may be more due to fluctuations in the device performance rather than advantage from the code. Nevertheless, even on this small-scale circuit we observe break-even performance from the Iceberg code.

\begin{table}[tb]
    \centering
    \captionsetup{justification=centering}
    \caption{\textsc{Fidelity of \\``Transversal-$H$ Bell State'' Circuits}}
    \begin{tabular}{|c|c|c|c|c|}
        \hline
        \makecell{Run\\ Number} & \makecell{Fault\\ Tolerance} & \makecell{Unencoded\\ Baseline} & \makecell{Encoded\\ Fidelity} & \makecell{Detected\\ Errors} \\
        \hline
        \hline
        \rowcolor{blue!20}
        Run 1 & No & 97.85\% & 99.24\% & 9.52\% \\
        \hline
        \rowcolor{blue!20}
        Run 1 & Yes & 97.85\% & 98.70\% & 24.66\% \\
        \hline
        \rowcolor{blue!20}
        Run 2 & No & 99.02\% & 99.18\% & 11.18\% \\
        \hline
        Run 2 & Yes & 99.02\% & 98.66\% & 23.19\% \\
        \hline
        Run 3 & No & 98.73\% & 98.68\% & 11.23\% \\
        \hline
        \rowcolor{blue!20}
        Run 3 & Yes & 98.73\% & 99.08\% & 25.29\% \\
        \hline
    \end{tabular}
    \label{tab:transversal-bell-state}
\end{table}

The ``Transversal-$H$ Bell State'' circuit yielded above break-even in all three runs of the circuit, more than for any other experiment; twice with the non-fault-tolerant configuration, and twice with the fault-tolerant configuration. The key improvement this circuit has over the ``Bell State'' circuit is the use of the transversal $H$ operation, which is fault-tolerant in nature and greatly reduces the total gate overhead.

\begin{table}[tb]
    \centering
    \caption{\textsc{Fidelity of ``$XX$ Toffoli'' Circuits}}
    \begin{tabular}{|c|c|c|c|c|}
        \hline
        \makecell{Run\\ Number} & \makecell{Fault\\ Tolerance} & \makecell{Unencoded\\ Baseline} & \makecell{Encoded\\ Fidelity} & \makecell{Detected\\ Errors} \\
        \hline
        \hline
        Run 1 & No & 97.61\% & 93.46\% & 21.58\% \\
        \hline
        \rowcolor{blue!20}
        Run 1 & Yes & 97.61\% & 98.41\% & 90.77\% \\
        \hline
        Run 2 & No & 98.58\% & 95.98\% & 16.21\% \\
        \hline
        Run 2 & Yes & 98.58\% & 98.06\% & 77.29\% \\
        \hline
        Run 3 & No & 98.14\% & 96.47\% & 17.09\% \\
        \hline
        Run 3 & Yes & 98.14\% & 97.45\% & 77.00\% \\
        \hline
    \end{tabular}
    \label{tab:xx-toffoli}
    \vspace{-10pt}
\end{table}

For the ``$XX$ Toffoli'' circuit, exactly one of the runs surpassed the break-even point for the fault-tolerant configuration. This run was executed shortly after device recalibration, while the two runs which failed to show improvement were performed later in the device recalibration cycle. This suggests that proper calibration is important to the success of this routine.

Additionally, across all runs of the ``$XX$ Toffoli'' circuit, the fault-tolerant configuration consistently outperformed the non-fault-tolerant configuration. That said, the rejection rate for the fault-tolerant configuration was $\sim60\%-70\%$ higher than the non-fault-tolerant configuration. This is expected (the fault-tolerant configuration has many more error flags) and it indicates the sacrifice that is paid for advantage from a QED code---increased shot overhead. That is, if an exact number of shots are required from a circuit, then when adding a QED code one must also increase the number of shots (perhaps by many times) to account for the anticipated rejection rate.

% The difference between runs likely resulted from experiments being run several times throughout the course of a week. Thus, the calibration and performance of the device differed for each run. In the experiment surpassing break-even from applying error detection, the device performed the worst in the unencoded baseline across all experiments. This resulted in the highest amount of detected errors across all experiments as well. It is likely that the XX-Toffoli circuit, the device, performs operations with a fidelity very close to the fidelity needed to achieve break-even. Depending on the device performance at the time of experiment (i.e., T1 time, T2 time, gate fidelities), the device can possibly achieve break-even performance.

\begin{table}[t]
    \centering
    \caption{\textsc{Fidelity of ``$HH$ Toffoli'' Circuits}}
    \begin{tabular}{|c|c|c|c|c|}
        \hline
        \makecell{Run\\ Number} & \makecell{Fault\\ Tolerance} & \makecell{Unencoded\\ Baseline} & \makecell{Encoded\\ Fidelity} & \makecell{Detected\\ Errors} \\
        \hline
        \hline
        Run 1 & No & 98.44\% & 95.90\% & 16.60\% \\
        \hline
        Run 1 & Yes & 98.44\% & 97.59\% & 77.73\% \\
        \hline
        Run  2 & No & 98.05\% & 96.15\% & 17.63\% \\
        \hline
        Run  2 & Yes & 98.05\% & 95.22\% & 78.56\% \\
        \hline
        Run 3 & No & 98.19\% & 96.85\% & 17.77\% \\
        \hline
        Run 3 & Yes & 98.19\% & 96.47\% & 76.46\% \\
        \hline
    \end{tabular}
    \label{tab:hh-toffoli}
\end{table}

For the ``$HH$ Toffoli'' circuit, none of the runs achieved an encoded fidelity higher than the unencoded baseline. This is not surprising, as all configurations of this circuit are the most complex of their respective class. For the ``Transversal-$H$ Toffoli'' circuit, however, above break-even performance was achieved above on \textit{all} runs for \textit{only} the fault-tolerant configuration. Thus, across all points in the device calibration life cycle, the fault-tolerant configuration yielded a fidelity gain inaccessible to the non-fault-tolerant configuration. Once again, the fault-tolerant configuration achieved its advantage at the cost of a rejection rate $\sim60\%-70\%$ above that of the non-fault-tolerant configuration.

\begin{table}[tb]
    \centering
    \captionsetup{justification=centering}
    \caption{\textsc{Fidelity of \\ ``Transversal-$H$ Toffoli'' Circuits}}
    \begin{tabular}{|c|c|c|c|c|}
        \hline
        \makecell{Run\\ Number} & \makecell{Fault\\ Tolerance} & \makecell{Unencoded\\ Baseline} & \makecell{Encoded\\ Fidelity} & \makecell{Detected\\ Errors} \\
        \hline
        \hline
        Run 1 & No & 98.54\% & 97.34\% & 19.14\% \\
        \hline
        \rowcolor{blue!20}
        Run 1 & Yes & 98.54\% & 99.17\% & 82.28\% \\
        \hline
        Run  2 & No & 98.49\% & 97.85\% & 18.36\% \\
        \hline
        \rowcolor{blue!20}
        Run  2 & Yes & 98.49\% & 99.75\% & 80.57\% \\
        \hline
        Run 3 & No & 98.49\% & 95.07\% & 17.77\% \\
        \hline
        \rowcolor{blue!20}
        Run 3 & Yes & 98.49\% & 99.61\% & 87.45\% \\
        \hline
    \end{tabular}
    \label{tab:transversal-toffoli}
    \vspace{-10pt}
\end{table}

The results of the ``Transversal-$H$ Bell State'' and the ``Transversal-$H$ Toffoli'' circuits highlight an important point that can be overlooked when protecting a circuit with a QED or QEC code. The use of the transversal $H$ gates made for a more complex unencoded circuit, but a more efficient encoded circuit. Thus, circuits should be compiled prior to the addition of a QED or QEC code so as to create the leanest encoded circuit (e.g., code-optimal compilation). This is subtly different from optimizing the circuit before and after the addition of a QED or QEC code. The former technique allows for the compilation of an algorithm into a non-optimal circuit prior to the encoding, with the understanding that the encoded circuit will naturally be optimal. The latter technique entails compiling into an optimal circuit prior to encoding, which results in an non-optimal encoded circuit (i.e., even if the encoded circuit is reoptimized, it will be more complex than the circuit created by the former technique).

% Naively, one might expect the XX-Toffoli circuit to have the highest fidelities of all experiments with the Toffoli gates. The logical circuit performs operations only in the computational basis, thus removing the potential of errors occurring within the phase of a qubit. However, within the Iceberg code and under the same device restrictions, the H-Transversal circuit performs better. This is due to the H-Transversal gate. Since this operation is fault-tolerant in nature and acts on all qubits, it removes the need for three of the targeted Hadamards (which are expensive operations to implement). This allows the circuit to perform all operations fault-tolerantly within the limited qubit availability.

A bit more can be said about this for the Iceberg code specifically. When one is compiling an unencoded circuit, the two-qubit gate count is typically the dominant source of controllable error.\footnote{For some quantum computers measurement is more error-prone than two-qubit gate operations, however, the number of measurements in a circuit is often fixed, while the number of two-qubit gates is variable.} For this reason, unencoded circuits are compiled so as to minimize the number of two qubit gates. However, when considering the operations of the Iceberg code, the targeted $H$ gate becomes a very expensive operation requiring many physical gates, particularly when it is performed fault-tolerantly. The transversal $H$ operation, however, requires significantly fewer gates.\footnote{In fact, this operation requires no more gates than a transversal $H$ operation on a set of \textit{unencoded} qubits.} As such, the goal of optimizing a circuit for the Iceberg code must shift from minimizing the number of two-qubit gates to maximizing the opportunity for the alignment of $H$ gates into transversal columns.

% This operation also highlights how performing the same logical circuit with well-chosen operations within a codespace can improve performance. The HH-Toffoli and H-Transversal circuits are logically equivalent. However, the H-Transversal circuit was the only one of the two to perform better than break-even (and did so on every run). Circuit compilation within an encoded space is an important aspect of quantum error correction and detection, and can be the difference between achieving break-even performance or not. For physical circuits, it is often best to compile such that we minimize the number of two-qubit operations. However, for encoded circuits, performing the transversal Hadamard gate turned out to be more important. Comparing the H-transversal to the XX-Toffoli circuit, we see that the H-Transversal circuit outperforms the XX-Toffoli circuit, even though the XX-Toffoli appears to be simpler. Although the physical circuit for the H-Transversal gate is more prone to errors (since the XX-Toffoli has no phase, thus minimizing phase errors), the H-Transversal circuit reduces the overhead significantly once compiled into the Iceberg code. In other words, optimized compiling before adding QEC is not the same as compiling for the most efficient QEC code—and the latter technique will likely produce the optimal circuit.

\section{Concluding Remarks}\label{conclusions}
The results shown in this work demonstrate the beginning of the transition to achievable fault-tolerant quantum computing. In particular, we have demonstrated beyond-break-even error suppression using the Iceberg quantum error detecting (QED) code on several algorithm circuits. This gain was demonstrated on six out of twelve Bell state preparation experiments for encoded circuits that used both fault-tolerant and non-fault-tolerant operations. Moreover, beyond-break-even error suppression was consistently demonstrated on a fault-tolerant Toffoli circuit that used a transversal $H$ operation; a gain that was not achievable for the same circuit when using non-fault-tolerant operations. Likewise, this gain was not achieved on different encoding of the same circuit which did not leverage the transversal $H$ routine. Moreover, beyond-break-even performance was observed on one fault-tolerant run of a circuit in which a Toffoli acted on a non-entangled state.

Where the advantage from the Iceberg QED code was most variable, the variation in performance was likely influenced by several factors. First, variation in the hardware noise at different times within the device recalibration cycle likely affected our results. Additionally, the number of qubits available in the hardware impacted our fault-tolerant circuits; some operations in these circuits were constructed without fault-tolerance to keep the total circuit size within the device's capacity.\footnote{This issue could be alleviated with mid-circuit measurement support, the addition of more qubits on the device, or a combination of both.}

In spite of all these limitations, our results demonstrate that the Iceberg QED code achieved near or above the break-even error detection for small proof-of-concept circuits of high practical utility. Moreover, all but one encoded circuit experienced improved fidelity with fault-tolerant operations as compared to their non-fault-tolerant counterparts.

This work could be continued in multiple directions. First, it would be desirable to repeat experiments on a trapped-ion device with support for mid-circuit measurement. This would allow for the most compact representation of the Iceberg code's fault-tolerant operations. In that connection, since some fault-tolerance was sacrificed in order to fit the circuits within the number of qubits of the trapped-ion hardware, mid-circuit measurement would allow for full fault-tolerant versions of all circuits studied herein. Alternatively, performing these experiments on a device with larger qubit would allow for the same exploration. Second, the Toffoli gate is a building block for more complex algorithms (e.g., Grover's algorithm \cite{Robertson2025Grovers}). Increasing the size of the logical circuit to encapsulate larger subroutines or even logical circuits should be explored.

\section{Acknowledgments}
The authors thank the Amazon Web Services (AWS) team for their generous support in time and access to Amazon Braket and other compute resources.

\bibliographystyle{IEEEtran}
\bibliography{main}

\end{document}